\documentstyle[epsfig]{mn}

\input{psfig.sty}

\ifnfsstwo

\fi

\ifnfssone
  \newmathalphabet{\mathit}
    \addtoversion{normal}{\mathit}{cmr}{m}{it}
    \addtoversion{bold}{\mathit}{cmr}{bx}{it}

\fi

\ifoldfss

\fi

\loadboldmathitalic
\loadboldgreek

\title[Metallicity Distribution of Stars of dSph]
       {The Predicted Metallicity Distribution of Stars in Dwarf Spheroidal
Galaxies}
\author[G. A. Lanfranchi and F. Matteucci]
 {Gustavo A. Lanfranchi $^1$ and Francesca Matteucci$^2$\\
$^1$IAG-USP,
 R. do Mat\~ao 1226, Cidade Universit\'aria, 05508-900 S\~ao Paulo, 
SP, Brazil\\
 $^2$ Dipartimento di Astronomia-Universit\'a di Trieste,
  Via G. B. Tiepolo 11, 34131 Trieste, Italy}

\pubyear{2004}

\begin{document}
\maketitle

\begin{abstract}
We predict the metallicity distribution of stars and the 
age-metallicity relation for 6 Dwarf Spheroidal (dSph) galaxies 
of the Local Group by means of a chemical evolution model which 
is able to reproduce several observed abundance ratios and the 
present day total mass and gas content of these galaxies.
The model adopts up to date nucleosynthesis and takes into account 
the role played by supernovae of different types (II, Ia) allowing
us to follow in detail the evolution of several chemical elements (H, D, He, 
C, N, O, Mg, Si, S, Ca, and Fe). Each galaxy model is specified by 
the prescriptions of the star formation rate and by the galactic wind 
efficiency chosen to reproduce the main features of these galaxies. 
These quantities are constrained by the star formation histories of 
the galaxies as
inferred by the observed color-magnitude diagrams (CMD). The main 
conclusions are: i) five of the six dSphs galaxies are characterized 
by very low star formation efficiencies ($\nu = 0.005 - 0.5\; 
Gyr ^{-1}$)  with only Sagittarius having a higher
one ($\nu = 1.0 - 5.0\;Gyr ^{-1}$); ii) the wind rate is proportional to the 
star formation rate and the wind efficiency is high 
for all galaxies, in the range $w_i$ = 6 - 15; iii) a high wind 
efficiency is required in order to reproduce the abundance ratios
and the present day gas mass of the galaxies; iv) the predicted
age-metallicity relation implies that the stars of the dSphs
reach solar metallicities in a time-scale of the order of 2 - 6 Gyr, 
depending on the particular galaxy; v) the metallicity distributions 
of stars in dSphs exhibit a peak around [Fe/H] $\sim$ -1.8 to -1.5 
dex, with the exception of Sagittarius, which shows a peak around 
[Fe/H] $\sim$ 
-0.8 dex; iv) the predicted metallicity distributions of stars suggest
that the majority of stars in dSphs are formed in a range of metallicity
in agreement with the one of the observed stars.

\end{abstract}


\section{Introduction}

The metallicity distribution of stars and the age-metallicity 
relation (AMR) are very important constraints on chemical evolution 
models of all types of galaxies. The metallicity distribution 
of stars is representative of the chemical enrichment of the 
galaxy and provides information about the history of the 
chemical evolution and how it proceeded. The AMR, on the other 
side, provides information on  the evolution with time of the 
Fe abundance in the interstellar medium (ISM) and in the stars.  
These two features, coupled with the abundance 
ratios of chemical elements, impose strong constraints on 
chemical evolution models and limit the range of acceptable 
values for several model parameters. 
In fact, 
the AMR and the metallicity distribution are affected 
by the choices of the initial mass function (IMF), the star 
formation rate (SFR) and the infall of gas (Tosi 1988), whereas
the abundance ratios depend mainly on the adopted
nucleosynthesis, IMF and stellar lifetimes (Matteucci 1996). 

In the Milky Way, the metallicity distribution of stars in the 
disc is, generally, estimated through G- and K- 
dwarf stars (see Kotoneva et al. 2002). These 
stars warrant a homogeneous sample and, since they have
lifetimes equal to or greater than the age of the galaxy, they
can provide a complete picture of the chemical enrichment 
of the Galaxy. The first observations compared to simple 
models of chemical evolution of our galaxy led to the so-called
G- dwarf problem (van den Bergh 1962; Schmidt 1963), i.e. the 
number of metal-poor stars in the solar neighbourhood is 
lower than what is  expected from predictions of the simple 
model (see Tinsley 1980; Matteucci 1996 for reviews).
This problem has not been found either in 
the halo or in the bulge of the Galaxy (Carney et al. 1987;
Laird et al. 1988; Maciel 1999). The first observations 
revealed a distribution in the solar neighbourhood with 
almost no stars with [Fe/H]
lower than $\sim$ -1.0 dex and two tentative peaks around 
$\sim$ -0.4 dex and $\sim$ 0.1 dex (Pagel 1989; Sommer-Larsen
1991). More recent observations using up-to-date techniques
(Wyse $\&$ Gilmore 1995; Rocha-Pinto $\&$ Maciel 1996) and a more 
sophisticated chemical evolution model (Chiappini, Matteucci \& 
Gratton 1997 - hereafter CMG97) solved the problem. The new 
distributions showed the same lack of very metal-poor stars but,
contrary to the previous results, exhibited a well-defined peak
in metallicity between [Fe/H] = -0.3 and 0 dex. This peak is well
reproduced by the model of CMG97, which assumes a two-infall 
scenario for the formation of our Galaxy and correctly predicts a 
negligible number of stars with very low [Fe/H], in agreement 
with the most recent G - and K- dwarf metallicity 
distributions. 

In the bulge of the Milky Way, the metallicity distribution 
is estimated from oxygen abundance (given in the notation
(O/H = log(O/H) +12) observed in planetary nebulae (PN)
(Ratag et al. 1997; Cuisinier et al. 1999). In such a case, 
there is a similarity with the metallicity distribution 
of the stars of the solar neighbourhood. There
is a peak around (O/H)= 8.7, very few, if any, super metal
rich objects with supersolar abundances and very few 
metal poor objects. In order to make a better 
comparison with the
metallicity distribution of the disc stars, Maciel (1999)
converted the O/H abundances into [Fe/H] using the 
theoretical [O/H] vs. [Fe/H] relationship from Matteucci et al.
(1999) for both the galactic bulge and solar neighbourhood.
He concluded that the bulge distribution looks like the 
K giant distribution of the disc if the [O/H] vs. [Fe/H] relation
for the solar neighbourhood is adopted. However, theoretical 
studies (Matteucci \& Brocato 1990; Matteucci et al. 1999)
have predicted a quite different [O/Fe] vs. [Fe/H] relation for 
the Bulge relative to the solar neighbourhood.
In fact, they predicted that the majority of bulge stars
should show [$\alpha$/Fe]$>$ 0.

The age-metallicity relation is a poorer constraint on 
chemical evolution models for the Galaxy if compared to
the metallicity distribution, since it can be reproduced
very easily by various types of models. The extended data
derived by Edvardsson et al. (1993) from F stars reveals 
a considerable scatter at almost all ages which can be 
reproduced easily by any model of chemical evolution.
This scatter is, according to the authors, real and not 
due to observational uncertainties. In spite of that,
the AMR exhibits a general behaviour of increasing 
metallicity with increasing galactic age, with a slope
depending on the infall of gas, on the functional dependence of 
the SFR on the gas density and time, and on the adopted 
IMF (Tosi 1988; Sommer-Larsen $\&$ Yoshii 1990).

For Dwarf Spheroidal galaxies, unfortunately, 
neither the metallicity distribution of stars nor the AMR 
are available, making a comparison between model
predictions and observations not yet possible. 
Very recently, however, a tentative metallicity
distribution of stars was constructed for Sagittarius dSph
(Bonifacio et al. 2004), but the number of data points is
very low, allowing us to make only a qualitative comparison. 
In that 
sense, a true prediction
that shall be confirmed or refuted by future observations
can be obtained by means of chemical evolution
models which are able to reproduce other observational 
constraints. We use in this paper 
the model from Lanfranchi $\&$ 
Matteucci (2003) (hereafter LM03) for dSphs in order
to predict both the AMR and metallicity distribution of 
stars in dSph galaxies. It is worth noting that 
the LM03 model
is able to reproduce some observable features of dSphs,
such as several abundance ratios, present day total mass 
and present day gas mass fraction, by treating in details
the energetics and taking into account the yields of 
supernovae  of type II (SNe II) and Ia (SNe Ia) as well 
as the yields of intermediate massive stars (IMS). The LM03 
model explicited the very important role of the galactic 
wind on the evolution of these galaxies and 
its effects on the abundance ratios, SFR, and final 
total and gas masses. The wind is characterized by its
efficiency, i.e. the intensity of the rate of gas
loss, which is assumed in LM03 to have the same efficiency
for all galaxies.
Possible differences in the rate of gas loss among 
these galaxies are accounted for by the different star 
formation (SF) efficiencies adopted for each galaxy, 
because the rate of the wind is assumed to be 
proportional to the SFR. In spite of 
that, we adopt in this work
different wind efficiencies in order to get a better 
agreement with the observable data, namely the abundance
ratios and HI mass/total mass fractions and to study the effect of different
wind rates among galaxies. 
The other parameters of
the model such as SFR, IMF, number, duration and epochs of 
the episodes of SF, remain the same as the ones in LM03.
With this new approach, the AMR and the metallicity 
distribution of stars are predicted for a sample of 6
dSph galaxies of the Local Group and the influence of the 
galactic winds on these two features, as well as on the
evolution of these galaxies, are analysed here.

The paper is organized as follows: in Sect. 2 we present
the observational data concerning the dSph galaxies,
in Sect. 3 the adopted chemical evolution models and 
star formation prescriptions are described, in Sect. 4 
we describe the results of our models, and finally in Sect. 
5 we draw some conclusions. We use the solar abundances 
measured by Grevesse $\&$ Sauval (1998) when the chemical 
abundances are normalized to the solar values
([X/H] =  log(X/H) - log(X/H)$_{\odot}$).

\section{Data Sample}

The data sample collected here is the same as in LM03, 
but now we restrict  our study to only 6 galaxies of 
the Local Group which possess enough data to be compared to 
the predictions of the models. In that sense, we discard 
Fornax and Leo I and analise the data from Draco, 
Sagittarius, Sculptor, Sextan, Ursa Minor and Carina. 
For these galaxies the data consists of abundances of iron 
and some $\alpha$-elements such as O, Mg, Si and Ca obtained 
from the most recent high-resolution spectroscopy of red 
giant stars in these galaxies (Smecker-Hane $\&$ McWilliam 
1999; Bonifacio et al. 2000; Shetrone, Cot\'e $\&$ Sargent 
2001; Shetrone et al. 2003; Bonifacio et al. 2004). 
Despite the relative small number of data points, it is 
possible to compare the observed abundance ratios 
with the model predictions. 
Besides the abundance ratios, we compare the results of the 
models to other properties of the dSph galaxies taken from the 
review of Mateo (1998) and use the star formation histories 
inferred from color-magnitude diagrams (Hernandez, Gilmore 
$\&$ Valls-Gabaud 2000; Dolphin 2002) as constraints on the number,
epoch and duration of the episodes of SF. Unfortunately, there is
no available data regarding both the AMR and the metallicity
distribution of stars, with the exception of Sagittarius, for 
which there is an observed metallicity distribution, but 
with a very low number of data points. 
The predictions are, therefore, compared only
qualitatively to the abundances and ages of the observed stars.

\section{Models} 

In this work we use the chemical evolution model for dSphs
of LM03. The scenario representing the dSph galaxies is 
characterized by one or two long episodes of star formation 
(SF) and by the occurrence of very intense galactic winds. 
The model allows one to follow in detail the evolution of 
the abundances of several chemical elements, starting from 
the matter reprocessed by the stars and restored into the 
ISM by stellar winds and type II and Ia supernova explosions.

The main features of the model are:

\begin{itemize}

\item
one zone with instantaneous and complete mixing of gas inside
this zone;

\item
no instantaneous recycling approximation, i.e. the stellar 
lifetimes are taken into account;

\item
the evolution of several chemical elements (H, D, He, C, N, O, 
Mg, Si, S, Ca and Fe) is followed in detail;

\item
the nucleosynthesis prescriptions include the yields
of: Thielemann, Nomoto $\&$ Hashimoto (1996) and Nomoto et al. (1997) 
for massive stars
(M $> 10 M_{\odot}$), van den Hoeck $\&$ Groenewegen 
(1997) for low and intermediate mass stars 
($0.8 \le M/M_{\odot} \le 8$) and Nomoto et al. 
(1997) for type Ia supernovae.
The type Ia SN progenitors are assumed to be  white dwarfs in binary systems
according the formalism originally developed by Greggio \& Renzini (1983)

\end{itemize}

In our scenario, the dSph galaxies form through
a continuous infall of pristine gas until a mass of
$\sim 10^8 M_{\odot}$ is accumulated.  One crucial 
feature in the evolution of these galaxies is the occurrence 
of galactic winds, which develop when the thermal 
energy of the gas equals its binding energy (Matteucci $\&$
Tornamb\'e 1987). This quantity is strongly influenced by 
assumptions concerning the presence and distribution 
of dark matter (Matteucci 1992). A diffuse ($R_e/R_d$=0.1, 
where $R_e$ is the effective radius of the galaxy and $R_d$ is 
the radius of the dark matter core) but massive 
($M_{dark}/M_{Lum}=10$) dark halo has been assumed for each galaxy.

\subsection{Theoretical prescriptions} 

The basic equation describing the evolution in time
of the fractional mass of the element $i$ in the gas 
within a galaxy, $G_{i}$, is:

\begin{equation}
\dot{G_{i}}=-\psi(t)X_{i}(t) + R_{i}(t) + (\dot{G_{i}})_{inf} -
(\dot{G_{i}})_{out}
\end{equation}
where $G_{i}(t)=M_{g}(t)X_{i}(t)/M_{tot}$ is the gas mass in 
the form of an element $i$ normalized to a total fixed mass 
$M_{tot}$ and $G(t)= M_{g}(t)/M_{tot}$ is the total fractional 
mass of gas present in the galaxy at the time t. The quantity 
$X_{i}(t)=G_{i}(t)/G(t)$ represents the abundance by mass 
of an element $i$, with the summation over all elements 
in the gas mixture being equal to unity. $\psi(t)$ is the 
fractional amount of gas turning into stars per unit time, 
namely the SFR. $R_{i}(t)$ represents the returned fraction 
of matter in the form of an element $i$ that the stars 
eject into the ISM through stellar winds and supernova 
explosions; this term contains all the prescriptions 
concerning the stellar yields and the supernova progenitor 
models. The infall of external gas and
the galactic winds are accounted for by the two terms 
$(\dot{G_{i}})_{inf}$ and $(\dot{G_{i}})_{out}$, respectively.
The prescription adopted for the star formation history is the
main feature which characterizes the dSph galaxy models.

The SFR $\psi(t)$ has a simple form and is given by:

\begin{equation}
\psi(t) = \nu G(t)
\end{equation}
where $\nu$ is the efficiency of star formation, namely
the inverse of the typical time-scale for star formation,
and is expressed in $Gyr^{-1}$.

In order to get the best agreement with the abundance ratios,
total final mass, and final gas mass, $\nu$ is varied in 
each galaxy following the procedure and results of LM03.
The star formation is not halted even after the onset of the 
galactic wind but proceeds at a lower rate since a large
fraction of the gas ($\sim\, 10\%$) is carried out of the galaxy. 
The details of the star formation are given by the star 
formation history of each individual galaxy as inferred by 
CMDs taken from Dolphin (2002) and Hernandez, Gilmore 
$\&$ Valls-Gabaud (2000). It is adopted 1 or 2 episodes of
SF, with durations that vary from 3 Gyr to 13 Gyr 
(see Table 1 for more details).
 
The rate of gas infall is defined as:
\begin{eqnarray}
(\dot G_{i})_{inf}\,=\,Ae^{-t/ \tau}
\end{eqnarray}
with A being a suitable constant and $\tau$ the infall 
time-scale which is assumed to be 0.5 Gyr.

The rate of gas loss via galactic winds for each element 
{\it i} is assumed to be proportional to the star formation 
rate at the time {\it t}:

\begin{eqnarray}
(\dot{G_{i}})_{out}\,=\,w_{i} \, \psi(t)
\end{eqnarray}
where $w_{i}$ is a 
free parameter describing the efficiency of the galactic
wind. The wind is assumed to be differential, i.e. some 
elements, in particular the products of 
SNe Ia, are lost more efficiently than others from the galaxy 
(Recchi, Matteucci $\&$ D'Ercole 2001; Recchi et al. 2002). 
This fact translates into slightly different values 
for the $w_i$ corresponding to different elements. Here we will
always refer to the maximum value of $w_i$. The efficiency 
of the wind is different for each dSph galaxy.

The initial mass function (IMF) is usually assumed to be 
constant in space and time in all the models and is 
expressed by the formula:

\begin{equation}
\phi(m) = \phi_{0} m^{-(1+x)}
\end{equation}

where $\phi_{0}$ is a normalization constant.
Following LM03 we assume a Salpeter-like IMF (1955)
($x=1.35$) in the mass range $0.1-100 M_{\odot}$, but
we also compare the predicted metallicity distributions to
one from a model adopting a Scalo (1986) IMF.

In table 1 we summarize the adopted parameters for the models
of dSph galaxies.

\begin{table*}
\begin{center}\scriptsize  
\caption[]{Models for dSph galaxies. $M_{tot}^{initial}$ 
is the baryonic initial mass of the galaxy, $\nu$ is the star-formation 
efficiency, $w_i$ is the wind efficiency, and $n$, $t$ and $d$ 
are the number, time of occurrence and duration of the SF 
episodes, respectively.}
\begin{tabular}{lccccccc}  
\hline\hline\noalign{\smallskip}  
galaxy &$M_{tot}^{initial} (M_{\odot})$ &$\nu(Gyr^{-1})$ &$w_i$
&n &t($Gyr$) &d($Gyr$) &$IMF$\\    
\noalign{\smallskip}  
\hline
Sextan  &$5*10^{8}$ &0.01-0.3 &9-13 &1 &0 &8 &Salpeter\\
Sculptor &$5*10^{8}$ &0.05-0.5 &11-15 &1 &0 &7 &Salpeter\\
Sagittarius &$5*10^{8}$ &1.0-5.0 &9-13 &1 &0 &13 &Salpeter\\
Draco  &$5*10^{8}$ &0.005-0.1 &6-10 &1 &6 &4 &Salpeter\\
Ursa Minor &$5*10^{8}$ &0.05-0.5 &8-12 &1 &0 &3 &Salpeter\\
Carina &$5*10^{8}$ &0.02-0.4 &7-11 &2 &6/10 &3/3 &Salpeter\\
\hline\hline
\end{tabular}
\end{center}
\end{table*} 

\section{Results}

We compare the predictions from the chemical evolution models
described in the previous section with some quantities, 
such as abundance ratios ([O/Fe], [Si/Fe], [Mg/Fe], 
and [Ca/Fe]), total present day mass and total final gas mass, 
observed in a sample of 6 dSph galaxies from the Local
Group. For each galaxy a model was built separately, using as
a constraint for the SFR the SF histories inferred 
from CMDs (see Table 2). After defining the number, epoch 
and duration of the bursts (the other parameters are the 
same as in LM03), the SF efficiency and the wind efficiency 
were varied in order to best reproduce the observed abundance 
ratios and final total and gas masses (the model for each 
galaxy which gave the best agreement with the observed data 
was defined as the best model for that galaxy). The 
metallicity distribution of stars and the AMR were then 
computed by means of the best model for  each of the 6 galaxies. We 
adopted, as first choices for the SF and wind efficiencies
the values given in LM03: a SF efficiency in the range 
$\nu$ = 0.005 - 3 $Gyr^{-1}$ and a wind efficiency of 10. 
The difference is that, in LM03 the wind efficiency was 
varied only for the standard model and kept the same in 
the models for individual galaxies, whereas here 
it is varied together with the SF efficiency in 
all models.

\subsection{The [$\alpha$/Fe] ratio} 

\begin{figure}
\centering
\epsfig{file=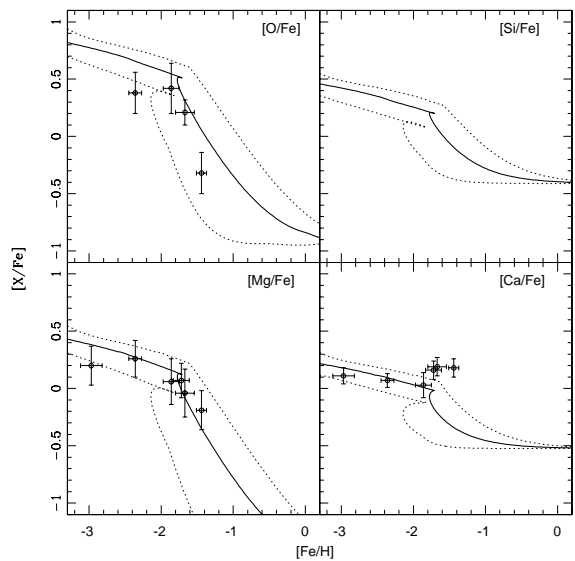,height=8cm,width=8cm}
\caption[]{[$\alpha$/Fe] vs. [Fe/H] observed in Draco dSph 
galaxy compared to the predictions of the chemical evolution
model for Draco. The thick solid line represent the best model 
($\nu = 0.03\;Gyr^{-1}$, w$_i$ = 6) and the dotted
lines the lower ($\nu = 0.005\;Gyr^{-1}$) and upper 
($\nu = 0.1\;Gyr^{-1}$) limits for the SF efficiency.}
\end{figure}

\begin{figure}
\centering
\epsfig{file=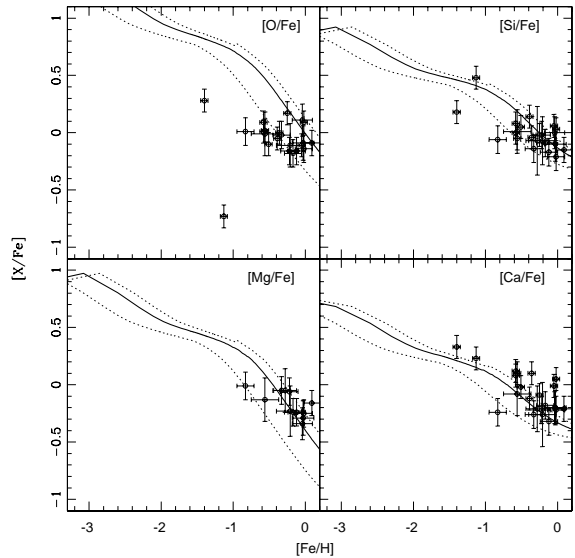,height=8cm,width=8cm}
\caption[]{[$\alpha$/Fe] vs. [Fe/H] observed in Sagittarius 
dSph galaxy compared to the predictions of the chemical evolution 
model for Sagittarius. The thick solid line represent the best 
model ($\nu = 3\;Gyr^{-1}$, w$_i$ = 9)
and the dotted lines the lower ($\nu = 1\;Gyr^{-1}$) and upper 
($\nu = 5\;Gyr^{-1}$) limits for the SF efficiency.}
\end{figure}

Abundance ratios are powerful tools in the study of chemical
evolution of galaxies because they depend mainly on the
nucleosynthesis prescriptions, stellar lifetimes
and adopted IMF and not on
the other model parameters. Some abundance ratios,
such as the [$\alpha$/Fe], can be used as ''chemical clocks", 
providing information about the SF time-scale due to the 
difference in the formation and injection of these elements 
into the ISM. The $\alpha$- elements are produced mainly in
SNe II explosions in a short time-scale while the Fe-peak
elements are produced in a much longer time-scale mainly in
explosions of SNe Ia. Consequently, a low [$\alpha$/Fe] implies
a long SF time-scale and an older age whereas a high [$\alpha$/Fe] ratio
is the result
of a short SF time-scaleand a younger age.

\begin{figure}
\centering
\epsfig{file=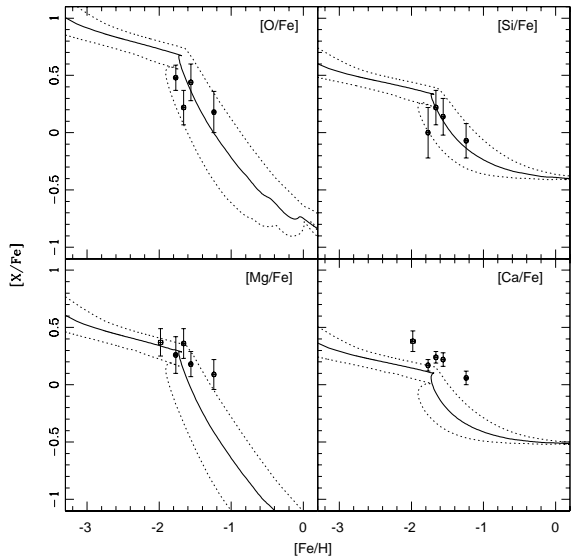,height=8cm,width=8cm}
\caption[]{[$\alpha$/Fe] vs. [Fe/H] observed in Sculptor dSph 
galaxy compared to the predictions of the chemical evolution 
model for Sculptor. The thick solid line represent the best 
model ($\nu = 0.2\;Gyr^{-1}$, w$_i$ = 13)
and the dotted lines the lower ($\nu = 0.05\;Gyr^{-1}$) and upper 
($\nu = 0.5\;Gyr^{-1}$) limits for the SF efficiency.}
\end{figure}

\begin{figure}
\centering
\epsfig{file=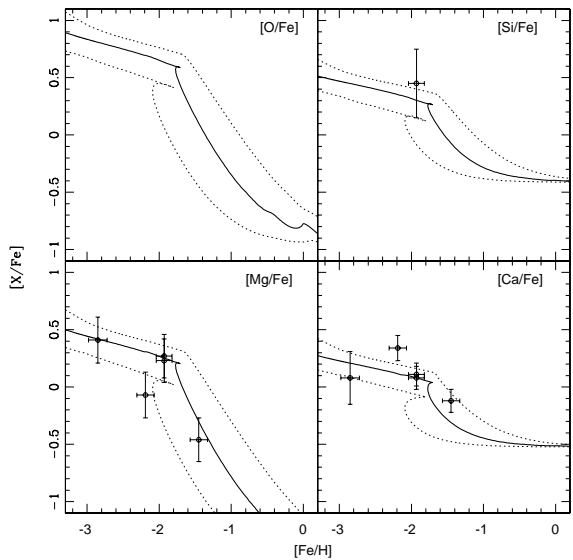,height=8cm,width=8cm}
\caption[]{[$\alpha$/Fe] vs. [Fe/H] observed in Sextan dSph 
galaxy compared to the predictions of the chemical evolution 
model for Sextan. The thick solid line represent the best model 
($\nu = 0.08\;Gyr^{-1}$, w$_i$ = 9)
and the dotted lines the lower ($\nu = 0.01\;Gyr^{-1}$) and upper 
($\nu = 0.3\;Gyr^{-1}$) limits for the SF efficiency.}
\end{figure}

The pattern of the [$\alpha$/Fe] 
observed in Local Group dSph galaxies, 
showing generally lower [$\alpha$/Fe] ratios
at the same [Fe/H] than in the Milky Way, 
is often considered difficult 
to explain (Tolstoy et al. 2003). This simply suggests that these
galaxies are characterized by a low SFR. LM03
found that 6 galaxies 
of the Local Group have [$\alpha$/Fe] ratios which can be well
reproduced by a model characterized by low SF efficiencies
in the range $\nu$ = 0.005 - 3 $Gyr^{-1}$, depending on the 
galaxy, and by the consideration of intense
differential galactic winds occurring in these galaxies.

We recover, in this work, the above analysis with a novelty
relative to LM03. We vary also the wind efficiency in
each galaxy in order to further explore the effects of 
the wind on the predicted abundance ratios and on the SF
efficiencies, restricting, therefore, the range of acceptable
values for this parameter. In Figures 1 to 6, the
[$\alpha$/Fe] ratios as functions of [Fe/H] are shown 
in comparison with the predictions of the chemical evolution models 
for each of the 6 dSph galaxies (Draco, Sagittarius, Sculptor, 
Sextan, Ursa Minor and Carina, respectively). The solid 
lines correspond to the best model for each galaxy while  the 
dashed lines represent the lower
and upper limits for the SF efficiencies. As one can see,
the models reproduce fairly well all the abundance ratios
in every galaxy, in the ranges of SF and wind efficiencies 
given by the models. Some galaxies do not have all 
abundance ratios available and, in that case, only the 
predictions are shown. The range of SF efficiencies is 
almost the same for four galaxies: Carina ($\nu 
= 0.02-0.4\;Gyr^{-1}$), Sculptor ($\nu = 0.05-0.5\;Gyr^{-1}$),
Ursa Minor ($\nu = 0.05-0.5\;Gyr^{-1}$) and Sextan
($\nu = 0.01-0.3\;Gyr^{-1}$). The other two galaxies
are reproduced by higher values of the SF 
efficiency - Sagittarius ($\nu  =\ 1.0-5.0\;Gyr^{-1}$) -
and lower ones - Draco ($\nu = 0.005-0.1\;Gyr^{-1}$).
The behaviour of the predicted abundance ratios is, though,
very similar in all cases and consistent with the trend
of the observed data: there is a sort of plateau 
in the abundance ratios for low metallicities 
([Fe/H] $\le$ $\sim$ -1.8, depending on the SF efficiency)
followed by a sharp decrease which is more pronounced
in the case of [O/Fe] and [Mg/Fe] than for
[Si/Fe] and [Ca/Fe]. 

\begin{figure}
\centering
\epsfig{file=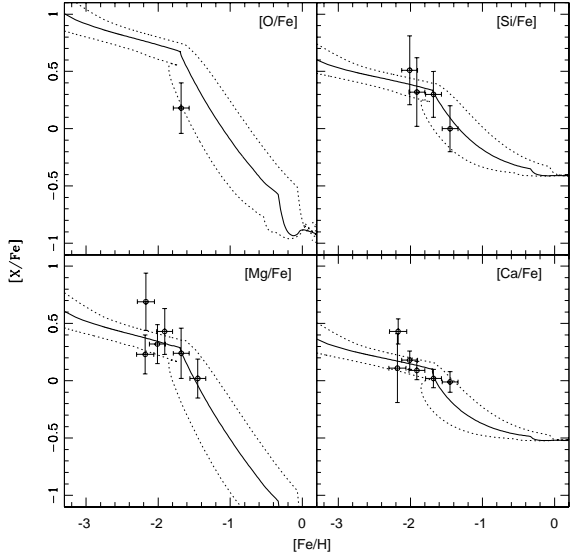,height=8cm,width=8cm}
\caption[]{[$\alpha$/Fe] vs. [Fe/H] observed in Ursa Minor dSph 
galaxy compared to the predictions of the chemical evolution model 
for Ursa Minor. The thick solid line represent the best model 
($\nu = 0.2\;Gyr^{-1}$, w$_i$ = 10)
and the dotted lines the lower ($\nu = 0.05\;Gyr^{-1}$) and upper 
($\nu = 0.5\;Gyr^{-1}$) limits for the SF efficiency.}
\end{figure}

\begin{figure}
\centering
\epsfig{file=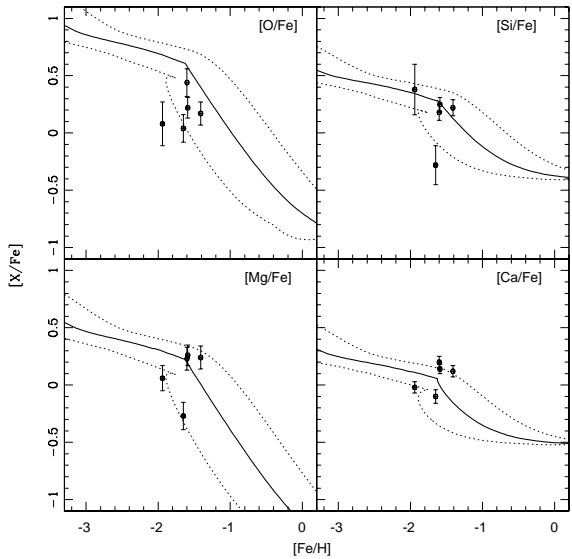,height=8cm,width=8cm}
\caption[]{[$\alpha$/Fe] vs. [Fe/H] observed in Carina dSph galaxy 
compared to the predictions of the chemical evolution model for Carina.
The thick solid line represent the best model ($\nu = 0.1\;Gyr^{-1}$,
w$_i$ = 7)
and the dotted lines the lower ($\nu = 0.02\;Gyr^{-1}$) and upper 
($\nu = 0.4\;Gyr^{-1}$) limits for the SF efficiency.}
\end{figure}

This sudden decrease in the predicted abundance ratios 
is a consequence of the occurrence of galactic winds and
of the injection of Fe in the ISM of the galaxy by SNe Ia.
As soon as the wind starts the SF declines fast due to the
decrease in the amount of the available gas almost halting
the formation of new stars and, consequently, the production
and injection of $\alpha$ elements into the ISM by SNe II.
The Fe-peak elements, on the other hand, are continuously
injected into the ISM by SNe Ia, since these explosions 
start occurring several Myr ($>$ 30 Myr) after the onset of the SF and 
continue, even if the SF is halted, for several Gyrs (up to 10) due to
the long lifetimes of the stars responsible for this type 
of explosion. These two facts are the main responsibles
for the general decrease of the [$\alpha$/Fe] ratios after
the onset of the wind. The differences between the pattern 
of the [Ca, Si/Fe] and that of [O,Mg/Fe] ratios reflect this 
behaviour.
There is a smoother decrease in [Ca/Fe] and [Si/Fe], 
since Ca and Si are produced also in SNe Ia and an 
abrupt one in [O/Fe] and [Mg/Fe], 
because the production of O and Mg in SNe Ia is negligible.

The effects of the differential aspect of the wind are 
clearly seen, though, only in the models with the lowest 
SF efficiencies and in the absolute Fe abundance (see LM03). 
The intensity of the wind, on the other hand, is very 
important to reproduce the abundance ratios, especially
the ones with the lowest values at higher metallicities. 
Generally, if the efficiency of the wind is increased, the 
decrease in the abundance ratios is more pronounced and 
the predicted values are lower after the onset of the wind.
This is caused by the influence of the wind on the SF: 
the more intense is the wind,  the lower is the SFR after 
the onset of the wind, since a higher fraction of gas 
($\sim 10\%$) is lost from the galaxy. In this sense, 
only an intense 
and differential wind can account for the lowest values 
observed in these galaxies. The intensity of the wind is
characterized by the wind efficiency, which lies in
similar ranges for all the dSph galaxies studied here,
close to the values defined for the models of LM03 (see Table 1).
The best model for each galaxy, on the other hand,
adopts wind efficiencies which differ considerably. While
the best model for Draco and Carina requires the lowest wind 
efficiencies ($w_i$ = 6 and 7, respectively) and for Sculptor
the highest one ($w_i$ = 13), the wind efficiency of the
best models for Ursa Minor, Sextan and Sagittarius assume
intermediate values ($w_i$ = 10, 9 and 9, respectively).
The differences in the best value and the similarity in 
the range of acceptable values of the wind efficiency
for all galaxies reflects the fact that each galaxy follows 
a particular
track of evolution, but they all resemble to each other, as
already noticed for the SF efficiency.

\begin{table*} 
\begin{center}\scriptsize  
\caption[]{Predictions of the models for dSph galaxies compared to 
observations. M$_{tot}^{final}$ is the present day total mass of the 
galaxy,  M$_{HI}$/M$_{tot}$ is the present day ratio between the HI
mass and total mass.}
\begin{tabular}{c|cc|cc|cc|cc}  
\hline\hline\noalign{\smallskip}  
   &\multicolumn{2}{|c|}{M$_{tot}^{final}$ (10$^6$ M$_o$)}  
&\multicolumn{2}{|c|}{M$_{HI}$/M$_{tot}$.10$^{-3}$}  
&\multicolumn{2}{|c|}{1$^o$burst (Gyr)} 
&\multicolumn{2}{|c|}{2$^o$burst (Gyr)}\\    
\noalign{\smallskip}  
\hline
 &obs &mod &obs &mod &obs &mod &obs &mod\\
\hline
Draco  &22$^a$ &5.53-25.5  &$<$1$^a$ &0.2-0.8 &6-10$^b$ &6-10 &- &-\\
Sagittarius &-- &70.4-214 &$<$1$^a$ &0.2-0.3 &0-13$^c$ &0-13 &- &-\\
Sextan &19$^a$ &6.19-28.9 &$<$1$^a$ &0.1-0.4 &- &0-8  &- &-\\
Sculptor  &6.4$^a$ &10.2-31.2 &4$^a$ &0.1-0.4 &0-7$^c$ &0-7 &- &-\\
Ursa Minor  &23$^a$  &11.3-38 &$<$2$^a$ &0.2-0.4  &0-3$^b$ &0-3 
&- &-\\
Carina &13$^a$  &9.72-58 &$<$1$^a$ &0.2-0.6 &6-9$^b$ &6-9 &10-13$^b$ 
&10-13\\
\hline\hline
\end{tabular}

a - Mateo (1998)

b - Hernandez, Gilmore $\&$ Valls-Gabaud (2000)

c - Dolphin (2002)

d - for the whole sample of dSph galaxies in Mateo (1998)
\end{center}
\end{table*} 

The total final mass and final gas mass observed are also
compared to the predictions of the models (see Table 2).
The ranges in the 
predicted values of the total mass and HI/total mass ratio
for each galaxy are a consequence of the variations in the SF and 
wind efficiencies. As the efficiency of the wind is increased the
total mass and final gas mass decrease, since more gas is lost
from the galaxy. Besides that, the total mass of stars is 
also lower since the decrease in the SFR after the onset of the 
wind is more pronounced for higher values of the wind efficiency. 
The SF efficiency, on the other hand, acts in 
an opposite way. If it is increased,  the final total mass and 
present day gas mass are also increased. These effects, however, 
are less important than the ones due to the different wind 
efficiencies. Consequently, the range of values adopted for 
the wind efficiency for each galaxy produces a range of predicted 
masses for each model galaxy.
In particular, the range of adopted values 
for the wind and SF efficiencies gives final gas masses 
in agreement with what is inferred from observations, 
suggesting that our assumptions on galactic winds and SFR 
are quite reasonable.

\subsection{The age-metallicity distribution}

\begin{figure}
\centering
\epsfig{file=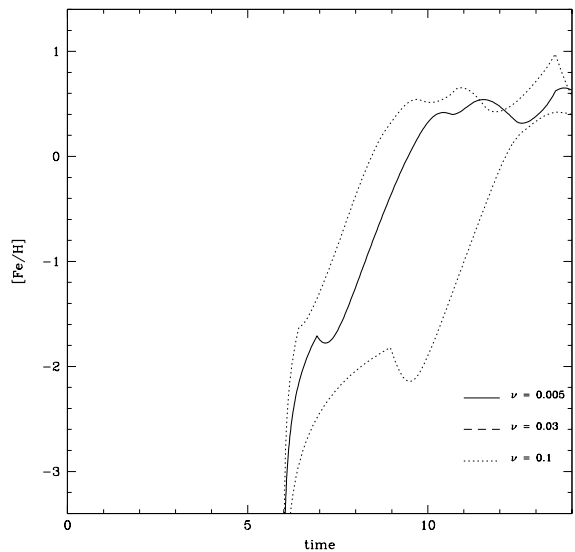,height=8cm,width=8cm}
\caption[]{[Fe/H] as a function of time as predicted by the model
for Draco. The solid line represents the best model and the dotted
lines the upper and lower limits on the SF efficiencies.}
\end{figure}

\begin{figure}
\centering
\epsfig{file=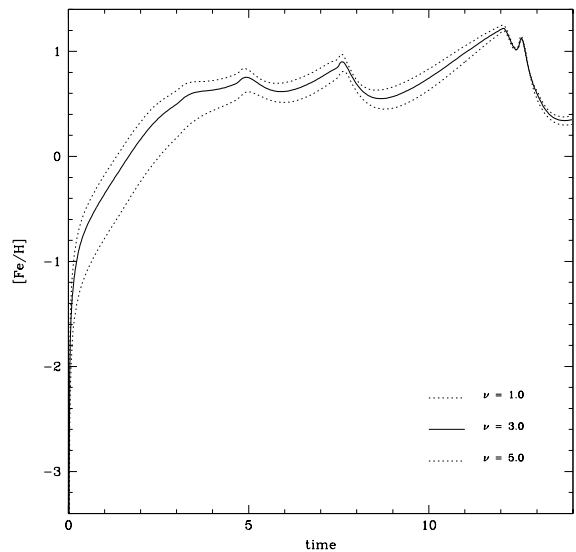,height=8cm,width=8cm}
\caption[]{[Fe/H] as a function of time as predicted by the model
for Sagittarius. The solid line represents the best model and 
the dotted lines the upper and lower limits on the SF 
efficiencies.}
\end{figure}

In Figures 7 to 12, the predicted age-metallicity relation is plotted 
for Draco, Sagittarius, Sculptor, Sextan, Ursa Minor and Carina, 
respectively. 
The solid lines represent the best model for 
each galaxy and the dotted lines the upper and lower limits on the 
SF efficiency. In general, the metallicity, represented by the 
predicted [Fe/H], increases with time reaching solar values
between 2 to 6 Gyr after the onset of the SF. This time-scale depends
on the details of the SF, such as the SF efficiency and the SF 
history of the galaxy. Concerning the SF history, the galaxies 
can be divided in to two distinct groups: one with galaxies in which
the SF starts at
the beginning of the formation (by gas assembly)
of the galaxy by gas accretion
and another one with
the galaxies in which the SF starts several Gyr after the formation 
of the galaxy. Carina and Draco are part of the second group and
the remaining galaxies, Sagittarius, Sextan, Sculptor and Ursa 
Minor, form the group of galaxies with SF occurring since the 
beginning of formation of the system.

\begin{figure}
\centering
\epsfig{file=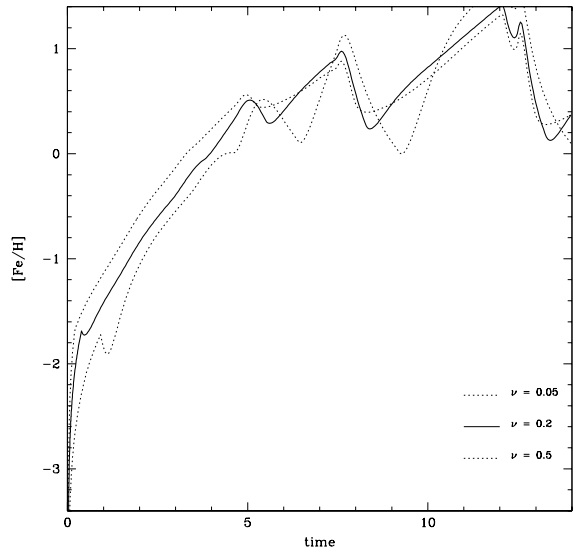,height=8cm,width=8cm}
\caption[]{[Fe/H] as a function of time as predicted by the model
for Sculptor. The solid line represents the best model and the 
dotted lines the upper and lower limits on the SF efficiencies.}
\end{figure}

\begin{figure}
\centering
\epsfig{file=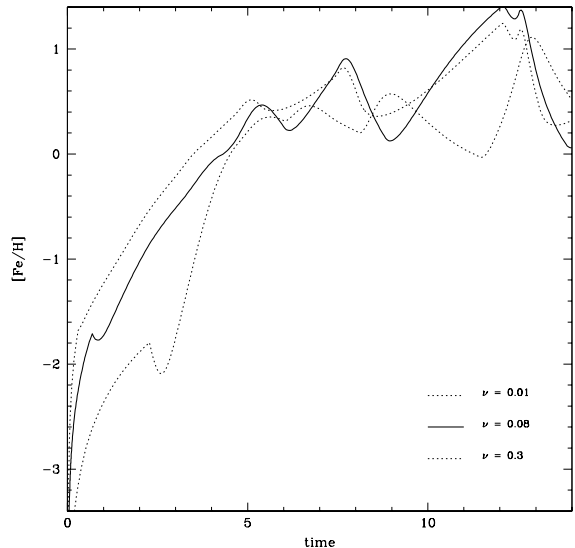,height=8cm,width=8cm}
\caption[]{[Fe/H] as a function of time as predicted by the model
for Sextan. The solid line represents the best model and the dotted
lines the upper and lower limits on the SF efficiencies.}
\end{figure}

The difference in the epoch of the onset of the SF between the two
groups reflects other major differences. In Draco and Carina 
the metallicity rises fast reaching solar values in a few Gyr, 
between 2.5 to 6 Gyr in Draco and 1.5 to 3.5 Gyr in Carina. In 
Carina the rise is faster due to the higher SF efficiency
($\nu$ = 0.02 - 0.4 Gyr $^{-1}$) compared to  Draco ($\nu$ 
= 0.005 - 0.1 Gyr $^{-1}$). This behaviour implies that the 
ages of the observed stars of these galaxies should be low, 
since they started forming around 8 - 9 Gyr ago and reached 
solar chemical composition  in the gas $\sim 2 - 3\;Gyr$ later, 
and that the spread
in their ages should be relatively small, $\sim$ 2 - 3 Gyr for the 
best model of each galaxy. The spread in the ages of the observed
stars is inferred considering that all the observed stars have [Fe/H]
lying in the range -3.0 to 0.0 dex and assuming the SF history
inferred by the observed CMDs. 
However, both the spread in gas abundances and the time-scale 
necessary to reach the solar metallicities strongly depend on the 
adopted SF efficiency.

If one considers the models with the lower limit on the SF 
efficiency the spread in the ages among the stars will be between 
($\sim 3.5$ to $\sim6\; Gyr$) and the metallicities
will take longer to reach solar values. On the other hand, if $\nu$
is larger, these time intervals will be shorter, the spread in 
the ages of the stars narrower and the dominant stellar 
population will be older. Tolstoy et al. (2003)
have derived age-metallicity relations for Carina and Sculptor.
However, a meaningful comparison 
between our age-metallicity relations and theirs is not possible
since we adopted SFHs different from the ones they adopt. 
In particular, there
is a discrepancy between our stellar ages and those
inferred by Tolstoy et al. (2003) for Carina. 
In our opinion, firm conclusions on this point cannot still be drawn,
because of the many uncertainties affecting the derivation of 
stellar ages. Anyway, what is interesting to note is that our assumed SFHs 
well
reproduce the abundance patterns ([$\alpha$/Fe] vs. [Fe/H]).

The other group of galaxies, the ones with a SF beginning as soon 
as the galaxy starts forming (Sculptor, Sextan, Ursa Minor and 
Sagittarius), exhibit a different behaviour. In this case,
the dependence of the AMR on the SF efficiency is weak and the
results for the best model apply also to the models with the 
lower and upper limits on the SF efficiency. The rise of the 
[Fe/H] with time is less steep and it is necessary $\sim 4\; Gyr$ 
for the ISM to reach solar metallicities. This implies a
larger range for the stellar ages in these galaxies,
between $\sim 3.5-4.5\; Gyr$ for Sextan and Ursa Minor and 
$\sim 3.5-4.75\;Gyr$ for Sculptor, in agreement with the ages 
derived observationally (Tolstoy et al. 2003). A fraction of the
stars would be also older 
than $\sim 10\; Gyr$ since they started forming around 14 Gyr ago 
(assuming that the Universe has 15 Gyr) and would have reached 
solar metallicities 4 Gyr later. Sagittarius, however, does not
follow this behaviour. Even though the SF in this galaxy starts
as soon as the galaxy forms, the rise in the [Fe/H] is very fast
($\sim 2\;Gyr$) due to the higher SF efficiency compared to 
the other dSph galaxies. Because of these two facts, the stars
of Sagittarius would be the oldest ones (with ages $\sim 12\;Gyr$) 
among the stars observed in dSph galaxies. The fact that the
stars of Sagittarius should be old may lead to the idea that
they should also be less enriched, but, as the SF is very 
efficient in this galaxy, the chemical enrichment of the ISM 
is very efficient too.
In this way the stars of Sagittarius, although old, are also metal rich.
In fact, it is meaningless to compare the metallicities of the 
stars of different galaxies according to their age, because each
galaxy has had a particular evolution with a particular SF, 
even though there are similarities
among the SF histories of the local dSph galaxies.

\begin{figure}
\centering
\epsfig{file=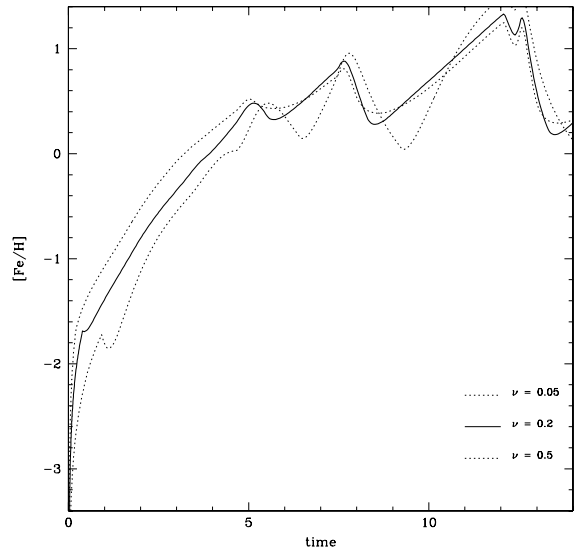,height=8cm,width=8cm}
\caption[]{[Fe/H] as a function of time as predicted by the model
for Ursa Minor. The solid line represents the best model and the 
dotted lines the upper and lower limits on the SF efficiencies.}
\end{figure}

\begin{figure}
\centering
\epsfig{file=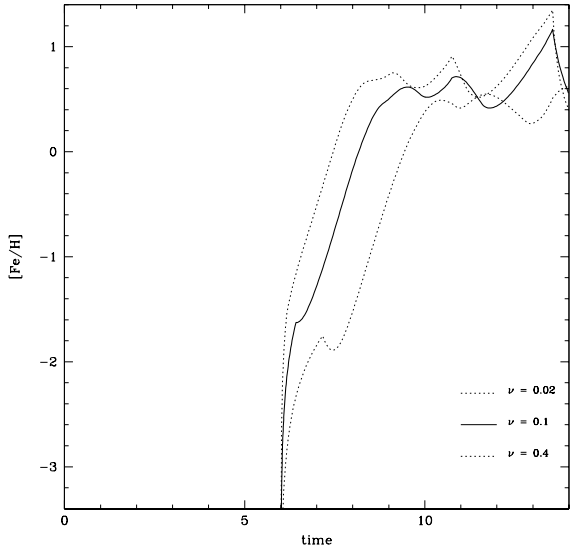,height=8cm,width=8cm}
\caption[]{[Fe/H] as a function of time as predicted by the model
for Carina. The solid line represents the best model and the dotted
lines the upper and lower limits on the SF efficiencies.}
\end{figure}

It is worth noting that all the model galaxies, 
independently of the adopted SF history or
SF efficiency, exhibit, however, peaks in the AMR. As the adopted
SF is characterized by long episodes ($>$ 3 Gyr) of activity one
should expect a continuous increase in metallicity, contrary
to what is predicted by the models. Actually, even though the
episodes of SF are long, the rate of SF is not constant over 
time and decreases substantially after the onset of the wind
due to the removal of a large fraction of the 
gas reservoir which fuels the SF. The formation of new stars 
is, consequently, almost halted and the injection of Fe into 
the ISM is negligible, except for the sporadic contribution of SNe 
Ia explosions. In this sense, the increase of the metallicity 
in the ISM is much less intense and, sometimes, there is 
even a decrease due to the preferential loss of Fe by 
means of differential galactic winds. The peaks seen in the 
predicted AMR are, then, a result of an interplay between
the injection of Fe into the ISM by SNe Ia explosions (when
the [Fe/H] increases) and the preferential loss of Fe by 
galactic winds (causing the decrease). This behaviour is 
responsible also for maintaining an almost constant mean 
metallicity several Gyr after the beginning of the SF.

\subsection{The stellar metallicity distribution}

The predicted metallicity distribution of stars of the Milky
Way (MW) disc at the solar neighbourhood compared to the 
ones predicted by the best model 
for Draco, Sagittarius, Sextan, Sculptor, Ursa Minor and Carina 
are shown in Figures 13 to 18, respectively. 
For Sagittarius the predicted metallicity distribution is 
compared also to the data of Smecker-Hane $\&$ Mc. William (1999) 
and Bonifacio et al. (2004). The solid lines 
represent the predictions for the best model selected for each 
dSph galaxy, the dotted line the predicted distribution 
of the model of CMG97 for the solar vicinity in the disc of 
the MW, which well reproduces the observed one, and the dashed line the
observational distribution for Sagittarius.

\begin{figure}
\centering
\epsfig{file=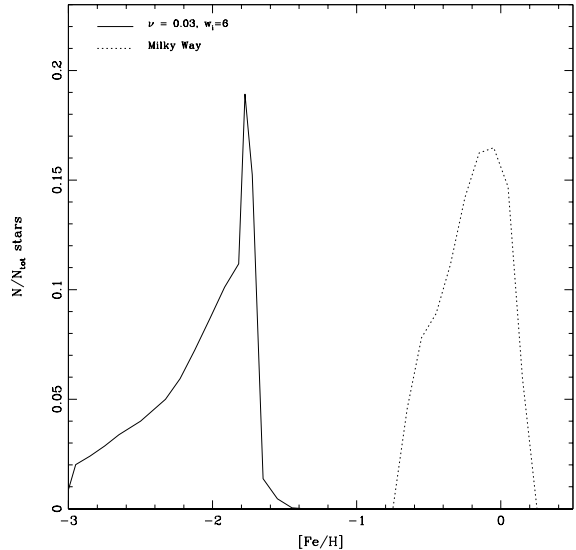,height=8cm,width=8cm}
\caption[]{The distribution of stars as a function of [Fe/H]
predicted by the best model for Draco compared to the 
distribution of the Milky Way model (dotted line).}
\end{figure}

\begin{figure}
\centering
\epsfig{file=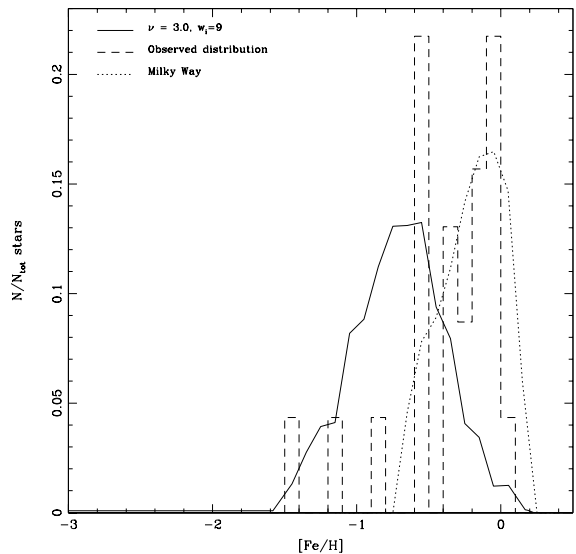,height=8cm,width=8cm}
\caption[]{The distribution of stars as a function of [Fe/H]
predicted by the best model for Sagittarius compared to the 
distribution of the Milky Way model (dotted line) and to the 
observed one (dashed line)(see text).}
\end{figure}

In general, the 
dSph galaxies exhibit distributions qualitatively similar to
the one predicted for the solar vicinity: there is a unique 
distinct peak in the relative number of stars formed at 
a given  metallicity. This is, however, the only similarity between
the distributions of the solar neighbourhood and
the dSph galaxies. The position in metallicity of the peak, its
width and the slope of the distribution function are different, 
not only between the dSph galaxies
and our Galaxy, but also between the dSphs themselves.

\begin{figure}
\centering
\epsfig{file=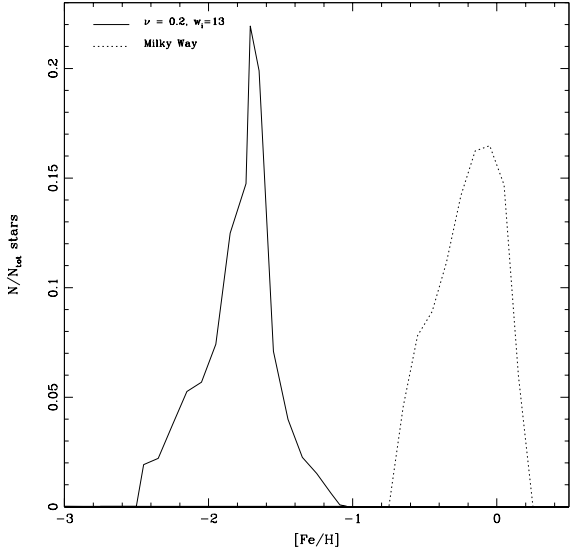,height=8cm,width=8cm}
\caption[]{The distribution of stars as a function of [Fe/H]
predicted by the best model for Sculptor compared to the 
distribution of the Milky Way model (dotted line).}
\end{figure}

\begin{figure}
\centering
\epsfig{file=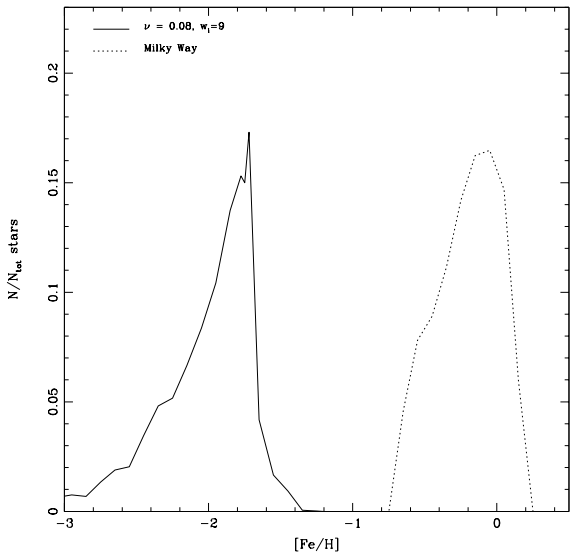,height=8cm,width=8cm}
\caption[]{The distribution of stars as a function of [Fe/H]
predicted by the best model for Sextan compared to the 
distribution of the Milky Way model (dotted line).}
\end{figure}

In general, the peak of the metallicity distribution in the 
dSphs occurs at lower metallicities than the one  
in the solar neighbourhood. The position of 
the peak, though, is specific for each galaxy.
While Sagittarius has a peak near the one of the Milky Way 
(at [Fe/H] $\sim$ -0.8 dex in Sagittarius and at [Fe/H] 
$\sim$ -0.1 dex in the MW), for Draco the peak is at [Fe/H] 
$\sim$ -1.8 dex. Generally, in the dSphs the peak is situated
around $\sim$ -1.6 dex, almost 1.5 dex below the value of 
the solar vicinity, but close to the value inferred for the halo 
metallicity distribution, which lies around [Fe/H] 
$\sim$ -1.55 dex (Laird et al. 1988). The difference between
the peak of the dSph metallicity distributions and the one
of the solar vicinity is due to the fact that the dSph galaxies 
exhibit a lower SF efficiency than the MW disc, so the typical 
metallicity of the formed stars is lower than the one of the 
disc of our galaxy at the solar vicinity. Besides that, the 
occurrence of the intense galactic wind prevents the 
dSph galaxies to form stars 
with relatively high metallicities, since it carries 
away a large fraction of the gas of the galaxy, thus  decreasing
the SF which almost stops shortly after the 
time at which the wind develops (at [Fe/H] $\sim$ -1.4 - 
-1.2 dex). This fact is more 
clearly seen in the high metallicity end of the distributions.
One can easily notice for Draco, Carina, Sculptor and Sextan a 
sudden and intense decrease in the metallicity distribution 
after the peak. This is the effect of the intense wind on the 
number of stars formed: the SF does not cease completely, but 
decreases substantially, thus decreasing also the number of stars
formed at metallicities larger than the ISM metallicity at the 
time of the wind.
The exact value of this metallicity
depends strongly on the SF efficiency but weakly on the
wind efficiency. 
On the other hand, the efficiency
of the wind is very important in predicting the decrease
of the distribution after the peak: the higher is the wind
efficiency, the more intense is the decrease and the lower
is the number of stars formed after that.

\begin{figure}
\centering
\epsfig{file=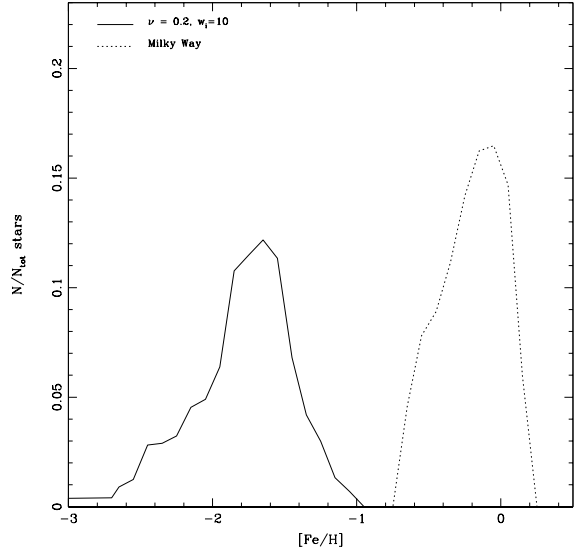,height=8cm,width=8cm}
\caption[]{The distribution of stars as a function of [Fe/H]
predicted by the best model for Ursa Minor compared to the 
distribution of the Milky Way model (dotted line).}
\end{figure}

\begin{figure}
\centering
\epsfig{file=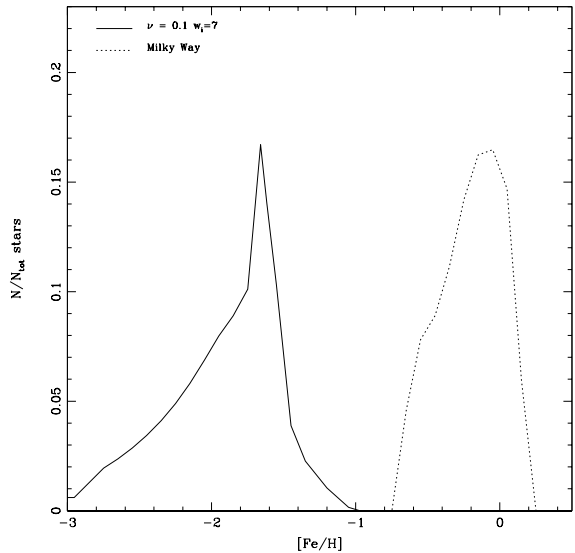,height=8cm,width=8cm}
\caption[]{The distribution of stars as a function of [Fe/H]
predicted by the best model for Carina compared to the 
distribution of the Milky Way model (dotted line).}
\end{figure}

The extension toward low metallicities of the predicted metallicity 
distribution in dSph galaxies exhibits a different pattern than
the one of our Galaxy. In the solar neighbourhood, there are almost
no stars formed with metallicities lower than $\sim$ -1.0 dex
(the so called G-dwarf problem) contrary to the dSph galaxies 
where there are stars with metallicities well below the one of
the peak. This is due to the long infall time-scale required by 
the metallicity distribution in the solar vicinity ($\sim 7-8\;$ Gyrs)
as opposed to the short infall time-scale assumed for dSphs (0-5 Gyrs).
The spread in the metallicity of the majority of stars
in dSph galaxies is, consequently, larger than in the solar neighbourhood.
While in the MW the metallicity distribution of
stars spans $\sim$ 1.0 dex in metallicity, in the dSph it reaches
$\sim$ 1.6 dex. This difference is related to the much lower SF
rate in the dSph galaxies when compared to the one of the disc
of the MW ($\nu=1 Gyr^{-1}$, see CMG97).

\begin{figure}
\centering
\epsfig{file=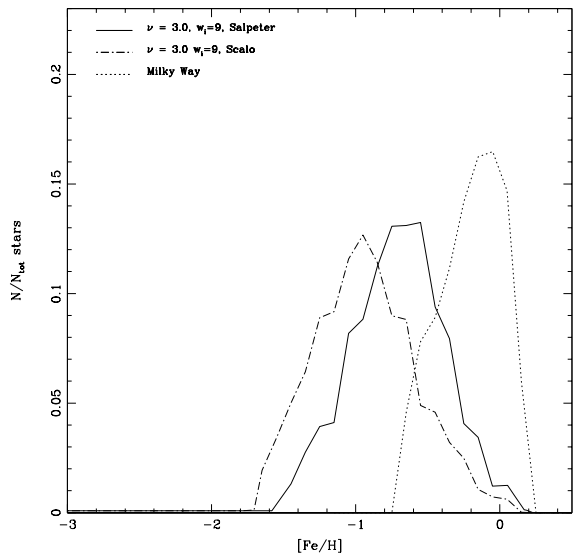,height=8cm,width=8cm}
\caption[]{The distribution of stars as a function of [Fe/H]
predicted by the model for Sagittarius with a Scalo IMF (dotted 
dashed line)and a Salpeter IMF (solid line) compared to the 
distribution of the Milky Way model (dotted line).}
\end{figure}

Even though there is no available data to make a comparison 
with the predicted distributions in dSph galaxies (with the 
exception of Sagittarius) it should be noticed that the
peaks for all galaxies lie in the same range of metallicities
of the observed stars in each galaxy. This suggests that the proposed
models are reasonable. In the case of Sagittarius, as the number
of the data points is very low, the comparison with the predictions 
is only qualitative. In that sense, the best model reproduces
very well the observed distribution, since the predicted range of 
metallicites of the formed stars is the same of the observed ones. 
The position of the peak of the observed distribution is not statistically
meaningful and should be confirmed or refuted by a more complete
data sample.

Finally, we computed a model similar to the one for Sagittarius but 
with a Scalo IMF in order to test the effect of a different IMF
and to compare more precisely
the predicted metallicity distribution of dSph galaxies with
the one of the model of CMG97 for the MW disc,
which adopts a Scalo IMF. When this
IMF is adopted for the dSph galaxies the predicted metallicity 
distribution is pushed toward lower values of metallicities and 
the peak of the distribution occurs at lower 
[Fe/H] values. Consequently, only the observed stars with the lowest
metallicities are fitted by the distributions predicted by
the model with Scalo IMF, that is unable to reproduce the stars 
with the highest metallicities. This fact together with the fact
that a model with Scalo IMF is unable also to reproduce the
abundance ratios (LM03) makes clear that this choice of IMF is
not suited for the dSph galaxies, which are better reproduced by
a Salpeter one.

\section{Summary}

We first analysed the star formation and chemical evolution in 6
dSph galaxies of the Local Group comparing several observed abundance 
ratios to the predictions of detailed chemical evolution models.
After defining the best models for each galaxy, which are specified 
by the SF history inferred by observed CMDs and by the adopted SF and 
wind efficiencies, we predicted the metallicity distribution of 
stars and the age-metallicity relation for each dSph of the sample.
By taking into account the role played by supernovae of 
different types (II, Ia) and adopting up to date nucleosynthesis
prescriptions we followed the evolution of several chemical elements 
(H, D, He, C, N, O, Mg, Si, S, Ca, and Fe). The predictions 
of the models were compared with the [$\alpha$/Fe] ratios and to the
present day total and gas masses. Since there are very few available data
concerning the metallicity distribution we compared our predictions 
to the MW disc distribution at the solar neighbourhood and to 
the metallicities of the observed stars. The observed
stars were also used to infer if the predicted age-metallicity
relations are reasonable, since there is also not enough data to
derive this relation observationally.

The main conclusions can be summarized as follows: 
\begin{itemize}

\item
five of the six dSph galaxies are characterized by very low 
star formation efficiencies ($\nu = 0.005 - 0.5\; Gyr ^{-1}$) 
with only Sagittarius having a higher one ($\nu = 1.0 - 5.0 \;Gyr^{-1}$).
In fact, Sagittarius is a unique galaxy when compared to the 
other dSphs of the sample. It is characterized by a much higher 
star formation efficiency. We predict for this galaxy
a peak in the metallicity 
distribution well above (at [Fe/H] $\sim$ -0.8 dex)
the peaks in the other dSphs (at [Fe/H] $\sim$ -1.6 - -1.2 dex). 
The AMR of this galaxy 
increases faster than the other dSphs with similar
SF histories, reaching solar values shortly after 2 - 3 Gyr
from the onset of the SF;

\item 
a high wind efficiency, in the range $w_i$ = 6 - 15, is 
required in order to reproduce the [$\alpha$/Fe] ratios and 
the present day mass in the gas of the galaxies. 
The predicted [$\alpha$/Fe] vs. [Fe/H] plots show a rather 
short plateau at low metallicities until a sharp decrease occurs 
at higher metallicities. This sudden 
decrease in the abundance ratios is a consequence of the
occurrence of intense galactic winds and of the SNe Ia explosions.
As soon as the wind starts, the SF decreases substantially 
and the rate of SNe II goes almost to zero halting
the injection of $\alpha$ elements into the ISM, whereas SNe 
Ia continue producing and injecting iron peak elements into
the ISM. These two facts cause the drop of the 
[$\alpha$/Fe] ratio;

\item 
the predicted age-metallicity relations imply that the stars 
of the dSphs reach solar metallicities on a time-scale 
of the order of 2 - 6 Gyr, depending on the particular 
SF history of the galaxy.
The stars of the galaxies with SF starting at the same time of 
the formation of the galaxy (Sagittarius, Sextan, Sculptor and
Ursa Minor) reach solar metallicities in a longer time-scale 
(tipically 4 - 5 Gyr) and are characterized by a wider
range of ages than the other two galaxies which have a SF starting
at a galactic age of 6 Gyr. These two galaxies, Carina and Draco, 
produce stars with solar metallicities 2 to 3 Gyr after the
beginning of the SF;

\item 
the metallicity distribution of stars of dSphs
exhibit a peak around [Fe/H] $\sim$ -1.8 to -1.5 dex, with 
the exception of Sagittarius, which shows a peak 
around [Fe/H] $\sim$ -0.8 dex. The typical peak is almost 
1.5 dex below the one of the MW disc in the solar 
neighbourhood but similar to the
value inferred for the halo of our Galaxy. The difference
in the predicted peaks of the dSph galaxies and the MW disc 
is a consequence of the fact that the dSph galaxies are 
characterized by much lower SF efficiencies and to the occurrence of
intense galactic winds which prevent the galaxies to form 
stars with metallicities higher than the one of the ISM of 
the galaxy shortly after the time at which the wind develops;

\item 
the predicted metallicity distributions of stars exhibit a peak
in the same range of metallicity of the observed stars, implying
that the majority of stars are formed in the same range of the
observed ones, if a Salpeter IMF is adopted. A Scalo-like IMF, as 
adopted in most models for the MW, predicts peaks at too low 
metallicities if compared with the available observations.

\end{itemize}

\section*{Acknowledgments}
We thank Antonio Pipino and Cristina Chiappini for reading the 
manuscript and for the comments on the paper.
G.A.L. acknowledges financial support from the Brazilian agency 
FAPESP (proj. 00/10972-0). F.M. acknowledges financial support  
from INAF Project ``Blu Compact Galaxies: primordial helium and chemical evolution'' and from COFIN2003 from the Italian Ministry for Scientific 
Reasearch (MIUR) project ``Chemical Evolution of Galaxies: interpretation 
of abundances in galaxies and in high-redshift objects''.

\bsp

\end{document}